\documentclass[11pt,twoside,a4paper,cmspaper,final,collab]{cms-tdr}
\def\svnVersion{9816472}\def\svnDate{2026/06/04}\def\cmsCernNoTag{CERN-EP-2025-196}\def\cmsCernDate{\today}\def\cmsMessage{Published in Physical Review D as \href{http://dx.doi.org/10.1103/mw5c-5rw3}{\doi{10.1103/mw5c-5rw3}.}}
\begin{document}\cmsNoteHeader{BPH-23-001}

\newcommand{\rjpsi}{\ensuremath{\mathcal{R}_\JPsi}\xspace}
\newcommand{\rjpsihad}{\ensuremath{\mathcal{R}_\JPsi^{\text{had}}}\xspace}
\newcommand{\rjpsihadeighteen}{\ensuremath{\mathcal{R}_\JPsi^{\text{had,\,2018}}}\xspace}
\newcommand{\result}{\ensuremath{0.49 \pm 0.11\stat \pm 0.23\syst \pm 0.07\thy}\xspace}
\newcommand{\resultc}{\ensuremath{0.49 \pm 0.26}\xspace}
\newcommand{\theoryresult}{\ensuremath{0.258 \pm 0.004}\xspace}
\newcommand{\bctojpsi}{\ensuremath{\PBpc \!\to\! \JPsi}\xspace}
\newcommand{\bcjpsiX}{\ensuremath{\bctojpsi \PX}\xspace}
\newcommand{\pp}{\ensuremath{\Pp\Pp}\xspace}
\newcommand{\BCVEGPY}{\texttt{BCVEGPY}}
\newcommand{\BcsignalMuon}{\ensuremath{\bctojpsi\PGmp\PGnGm}\xspace}
\newcommand{\BcsignalTau}{\ensuremath{\bctojpsi\PGtp\PGnGt}\xspace}
\newcommand{\RJpsi}{\ensuremath{\rjpsi = \mathcal{B}(\BcsignalTau) / \mathcal{B}(\BcsignalMuon)}\xspace}
\newcommand{\BcsignalLepton}{\ensuremath{\bctojpsi\Pell\PGn}\xspace}
\newcommand{\BcJpsiPi}{\ensuremath{\bctojpsi\PGpp}\xspace}
\newcommand{\pipi}{\ensuremath{\PGpp\PGpm}\xspace}
\newcommand{\pipipi}{\ensuremath{\pipi\PGpp}\xspace}
\newcommand{\BcJpsiPiPiPi}{\ensuremath{\bctojpsi\pipipi}\xspace}
\newcommand{\BcJpsiDs}{\ensuremath{\bctojpsi\PD_\PQs^{(\ast)+}}\xspace}
\newcommand{\Hc}{\ensuremath{\HepParticle{H}{\PQc}{}}\xspace}
\newcommand{\Hb}{\ensuremath{\HepParticle{H}{\PQb}{}}\xspace}
\newcommand{\BcJpsiX}{\ensuremath{\bctojpsi\Hc\PX}\xspace}
\newcommand{\HbJpsiX}{\ensuremath{\Hb \!\to\! \JPsi\PX}\xspace}
\newcommand{\tauHad}{\ensuremath{\PGt_\text{had}^{+}}\xspace}
\newcommand{\tauMudecay}{\ensuremath{\PGtp \!\to\! \Pgmp\Pgngm\Pagngt}\xspace}
\newcommand{\mumu}{\ensuremath{\PGmp\PGmm}\xspace}
\newcommand{\jpsimumu}{\ensuremath{\JPsi \!\to\! \mumu}\xspace}
\newcommand{\rBc}{\ensuremath{r_{\PBpc}}\xspace}
\newcommand{\aresonance}{\ensuremath{\HepParticleResonanceFull{a}{1}{}{1260}{}{+}}\xspace}
\newcommand{\rhomass}{\ensuremath{\HepParticleResonance{\PGr}{770}{}{0}}\xspace}
\newcommand{\taualphapi}{\ensuremath{\PGtp \!\to\! \aresonance\PAGnGt}\xspace}
\newcommand{\rhoz}{\ensuremath{\PGr^0}\xspace}
\newcommand{\rhoo}{\ensuremath{\PGr_1}\xspace}
\newcommand{\rhot}{\ensuremath{\PGr_2}\xspace}
\newcommand{\atorho}{\ensuremath{\aresonance \!\to\! \rhomass (\to\! \pipi) \PGpp}\xspace}
\newcommand{\bplustojpsikplus}{\ensuremath{\PBp \!\to\! \JPsi\PKp}\xspace}
\newcommand{\bplustopsiprimekplus}{\ensuremath{\PBp \!\to\! \Pgy(\to\! \JPsi\pipi)\PKp}\xspace}
\newcommand{\fexti}{\ensuremath{f_{\text{ext}}(i)}\xspace}
\newcommand{\bdt}{\ensuremath{s_\mathrm{BDT}}\xspace}
\newcommand{\srlow}{\ensuremath{4.2}\xspace}
\newcommand{\srlowl}{\ensuremath{3.5}\xspace}
\newcommand{\sbhigh}{\ensuremath{3.5}\xspace}
\newcommand{\sbhighl}{\ensuremath{3.0}\xspace}

\newlength{\cmsFigSize}
\ifthenelse{\boolean{cms@external}}{\setlength{\cmsFigSize}{0.49\textwidth}}{\setlength{\cmsFigSize}{0.7\textwidth}}
\newlength\cmsTabSkip\setlength{\cmsTabSkip}{1ex}
\providecommand{\cmsTable}[1]{\resizebox{\textwidth}{!}{#1}}
\ifthenelse{\boolean{cms@external}}{\providecommand{\cmsLeft}{upper\xspace}}{\providecommand{\cmsLeft}{upper left\xspace}}
\ifthenelse{\boolean{cms@external}}{\providecommand{\cmsMiddle}{middle\xspace}}{\providecommand{\cmsMiddle}{upper right\xspace}}
\ifthenelse{\boolean{cms@external}}{\providecommand{\cmsRight}{lower\xspace}}{\providecommand{\cmsRight}{lower\xspace}}

\cmsNoteHeader{BPH-23-001}
\title{Measurement of the ratio of the \texorpdfstring{\BcsignalTau}{Bc+ to J/psi tau+ nu} and \texorpdfstring{\BcsignalMuon}{Bc+ to J/psi mu+ nu} branching fractions using three-prong \texorpdfstring{\PGt}{tau} lepton decays}

\date{\today}

\abstract{
The ratio between the \BcsignalTau and \BcsignalMuon branching fractions is measured using a data sample of proton-proton collisions collected by CMS at a center-of-mass energy of 13\TeV in the years 2016--2018 and corresponding to an integrated luminosity of 138\fbinv. The \JPsi meson is identified through its \jpsimumu decay and the tau lepton is reconstructed in the hadronic three-prong final state. The measured ratio of branching fractions in this tau decay mode, $\rjpsihad = 1.04_{-0.44}^{+0.50}$, is combined with the previous analysis based on the \tauMudecay leptonic decay channel, leading to $\rjpsi=\resultc$. As this result is consistent with the standard model prediction of \theoryresult, no evidence of lepton flavor universality violation is found.
}

\hypersetup{%
pdfauthor={CMS Collaboration},%
pdftitle={Measurement of the ratio of the Bc+ --> J/psi tau+ nu and Bc+ --> J/psi mu+ nu branching fractions using three-prong tau lepton decays},%
pdfsubject={CMS},%
pdfkeywords={B physics, lepton flavor universality, CMS}} 

\maketitle 

The standard model~(SM) provides an excellent description of physical processes up to the electroweak scale and was completed by the discovery of the Higgs boson~\cite{ATLAS:2012yve,CMS:2012qbp,CMS:2013btf} by the ATLAS and CMS Collaborations at the CERN LHC\@. The Yukawa couplings of the Higgs boson to fermions are proportional to the fermion masses and can contain complex phases that contribute to charge-conjugation and parity violation, particularly through the CKM matrix~\cite{PhysRevLett.10.531,Kobayashi:1973fv} in the quark sector. 
The Yukawa couplings are also the only source of lepton flavor universality~(LFU) violation in the SM, and current measurements of the couplings are consistent with SM predictions for the second and third lepton generations~\cite{CMS:2018fdh,CMS:2018nsn}.
As LFU is not explained by any underlying structure of the SM, it is natural to consider more elaborate mechanisms for LFU\@. Observing LFU violation in \PB meson decays would indicate physics beyond the SM.

To test LFU, the LHCb Collaboration measured \RJpsi, using \jpsimumu and \tauMudecay decays, which ensures both decay modes have the same visible final state (three muons). The result, $0.71\pm 0.17\stat\pm 0.18\syst$~\cite{LHCb:2017vlu}, exceeds the value predicted by the SM, \theoryresult~\cite{Harrison:2020nrv,Anisimov:1998uk,Ivanov:2006ni, Hernandez:2006gt, Azizi:2019aaf}, by about two standard deviations.
The CMS Collaboration subsequently measured $\rjpsi = 0.17^{+0.18}_{-0.17}\stat^{+0.21}_{-0.22}\syst^{+0.19}_{-0.18}\thy$~\cite{CMS:2023vgr} in the same final states, using data collected in 2018 corresponding to an integrated luminosity of 59.7\fbinv~\cite{CMS-PAS-LUM-18-002} (described as the ``leptonic \PGt analysis'' in this paper).

The measurement presented in this paper, also provided in a HEPData record~\cite{hepdata}, is based on a sample of proton-proton~(\pp) collisions at $\sqrt{s} = 13\TeV$ collected by CMS in 2016--2018 corresponding to an integrated luminosity of 138\fbinv~\cite{CMS-PAS-LUM-18-002,CMS-PAS-LUM-17-004,CMS:2021xjt}.
We reconstruct the decay \BcsignalTau with subsequent decays of the \JPsi meson to a pair of muons and the \PGt lepton to its hadronic three-prong final state, reconstructed from three charged pions, referred to as \tauHad. Besides the neutrino, additional (unreconstructed) pions may also be present in the decay. The data used in Ref.~\cite{CMS:2023vgr} to evaluate the \BcsignalMuon yield are also used in the denominator of \rjpsi for this measurement. The result is combined with that of the leptonic \PGt analysis. In this paper, the charge conjugate of a particle or its decay mode is also implied.

The CMS apparatus~\cite{CMS:2008xjf,CMS:2023gfb} is a multipurpose, nearly hermetic detector, designed to trigger on~\cite{CMS:2020cmk,CMS:2016ngn,CMS:2024aqx} and identify electrons, muons, photons, and (charged and neutral) hadrons~\cite{CMS:2020uim,CMS:2018rym,CMS:2014pgm}. A global ``particle-flow''~(PF) algorithm~\cite{CMS:2017yfk} aims to reconstruct all individual particles in an event, combining information provided by the silicon tracker and by the lead-tungstate crystal electromagnetic, and brass and scintillator hadron calorimeters, operating inside a 3.8\unit{T} superconducting solenoid, with data from the gas-ionization muon detectors embedded in the flux-return yoke outside the solenoid. Objects such as hadronically decaying \Pgt leptons~\cite{CMS:2018jrd} are built from the reconstructed particles.

For each data-taking year, a sample of simulated events containing \PBpc mesons is produced with the \BCVEGPY~\cite{Bcveg1,Bcveg} Monte Carlo~(MC) event generator, interfaced with \EVTGEN 1.6.0~\cite{Ryd:2005zz,berezhnoy2011bcnpimoduleanalysisbc} for decays of weakly decaying hadrons, and with \PYTHIA 8.240~\cite{Sjostrand:2014zea} for parton showering, hadronization, and underlying event modeling using the \textsc{CP5} tune~\cite{CMS:2019csb}. The events include the signal modes, \BcsignalTau and \BcsignalMuon, and other \BcJpsiX decays, where \Hc indicates a (possibly excited) charm meson and \PX indicates additional products. The \TAUOLA~1.1.5 program is used to simulate \PGt decays \cite{Davidson:2010rw,Chrzaszcz:2016fte}. Contributions from \Pgy, \Pcgci, and \Pcgcii feed-down decays to \JPsi mesons are also included.
The branching fractions of the decays are obtained from their measured values or theoretical predictions~\cite{ParticleDataGroup:2022pth,LHCb:2019tea,LHCb:2013kwl,ATLAS:2015jep,LHCb:2016djy}. Most of the contributions that satisfy our selection criteria come from \BcJpsiDs decays. Contributions from other \PBpc decays $(\PBpc \!\to\! \text{others})$ are primarily from $\bctojpsi\PD{}^{(\ast)+}$, $\bctojpsi\PDp \PK{}^{(\ast)0}$, and $\bctojpsi\PD{}^{(\ast)0} \PKp$ decays. 
To account for decays not included in the simulation, the $\PBpc\!\to\!\text{others}$ yields are scaled up with a factor calculated from an investigation of the effect in simulation of other expected but unobserved \PBpc decay modes: \BcJpsiX, where \Hc includes charm meson excited states above the $\PD{}^{(\ast)}$ and \PX includes up to two pions or kaons, either neutral or charged. 
The \BcsignalLepton ($\Pell = \PGm,\,\PGt$) decay was generated with an hadronic model using the Kiselev form factor parametrization~\cite{KISELEV2000473}, updated to the parametrization in Ref.~\cite{bgl:2019} by reweighting the generated events using the \textsc{Hammer}~\cite{hammer} framework.
The events are also reweighted to match the world-average \PBpc mean lifetime value of 0.510\unit{ps}~\cite{PhysRevD.107.052008}. 
We initially normalize the \PBpc MC samples to the data in a two step process. The normalization of the 2018 MC sample is taken from the leptonic \PGt analysis~\cite{CMS:2023vgr} by comparing the \BcsignalMuon yields in the measured and simulated event samples. To normalize the 2016 and 2017 MC samples, we reconstruct and select \BcJpsiPiPiPi candidates, fit the invariant mass distributions of data and MC to obtain the yields for each year, calculate the data/MC ratio of yields, calculate the ratio of the 2016 and 2017 results to the 2018 result, and multiply by the 2018 normalization value. 

A second (background) simulated sample of \HbJpsiX decay events, where \Hb indicates a ground state \PQb hadron other than \PBpc, and the \JPsi decays to a pair of muons, is generated by \PYTHIA with the ``SoftQCD:nonDiffractive'' process.  One \Hb hadron in each event is forced to decay in a mode with a \JPsi meson. 
The \HbJpsiX decays are simulated with \EVTGEN, including feed-down decays of heavier charmonium states. This sample is used to derive corrections in the estimation of the inclusive \HbJpsiX background.

The simulated samples include effects from additional \pp interactions in the same or nearby bunch crossings (pileup), generated with \PYTHIA, with a multiplicity distribution tuned to match the data.  Final-state photon radiation is modeled by \PHOTOS~3.61~\cite{Barberio:1993qi}. The interactions between particles and the CMS detector are simulated using \GEANTfour~\cite{Agostinelli:2002hh} and the output is processed through the standard CMS reconstruction software.

Events are collected with a trigger that requires a pair of oppositely-charged muons, each with transverse momentum $\pt > 4\GeV$. The two muons are used to form a \JPsi candidate, which must have vertex fit probability above 10\%, $\pt>6.9\GeV$, and invariant mass of 2.9--3.3\GeV. The displacement in the transverse plane between the beam axis and the dimuon vertex is required to be at least three times the uncertainty in the displacement, \ie, a significance above 3. An additional track with $\pt>1.2\GeV$ is required to be present, form a vertex with the \JPsi candidate muons, and be displaced with respect to the beam axis with a significance above 2.

Offline, the two muons are required to match the muons used in the trigger, satisfy the trigger requirements, pass a set of loose identification~(ID) criteria~\cite{CMS:2018rym}, and have an invariant mass of 2.95--3.25\GeV. If multiple \JPsi candidates are found in an event (around 1\% of the events), the highest \pt one is selected.
We define the primary vertex~(PV) as the \pp collision vertex with the shortest distance along the beam axis to the line defined by the \JPsi candidate momentum direction that intersects the dimuon vertex.
Scale factors are derived for each data-taking year and applied to the simulated events to match the data efficiencies. The trigger efficiency comes from events that pass a different trigger than used for the signal and the offline muon reconstruction and identification efficiency uses the tag-and-probe method with \JPsi events~\cite{CMS:2018rym}. Three trigger scale factors are applied, one for the \JPsi selection binned in muon \pt and pseudorapidity $(\abs{\eta})$, one for the \JPsi displacement binned in the significance of the displacement in the transverse plane between the beam axis and the dimuon vertex, and one for the hadron track binned in track \pt, $\abs{\eta}$, and impact parameter significance. The muon scale factor is applied to both muons and is binned in muon \pt and $\abs{\eta}$.

Triplets of PF charged hadrons are used to create \tauHad candidates, with each charged hadron required to have $\pt>0.5\GeV$, $\abs{\eta}<2.5$, pass within 1.2\unit{mm} of the PV in the longitudinal direction, have $\sqrt{\smash[b]{(\Delta\eta)^2 + (\Delta\phi)^2}}<1$, where $\Delta\eta$ and $\Delta\phi$ are measured with respect to the \JPsi candidate momentum, and a distance of closest approach with respect to the \JPsi vertex between $-0.4$ and 0.6\unit{mm}, where the sign depends on the angle between the track and the \JPsi momentum direction. The triplet must have a charge of $\pm$1, vertex fit probability above 10\%, $\pt>3\GeV$, displacement significance with respect to the PV above 3, and invariant mass below 1.7\GeV. At least one pion candidate must match the track object found by the trigger.
When multiple triplets are found, we select the highest-\pt one. This occurs in 20\% of simulated signal events and the selection is correct in 70\% of the events.

\begin{figure}[htb!]
\centering
\includegraphics[width=\cmsFigSize]{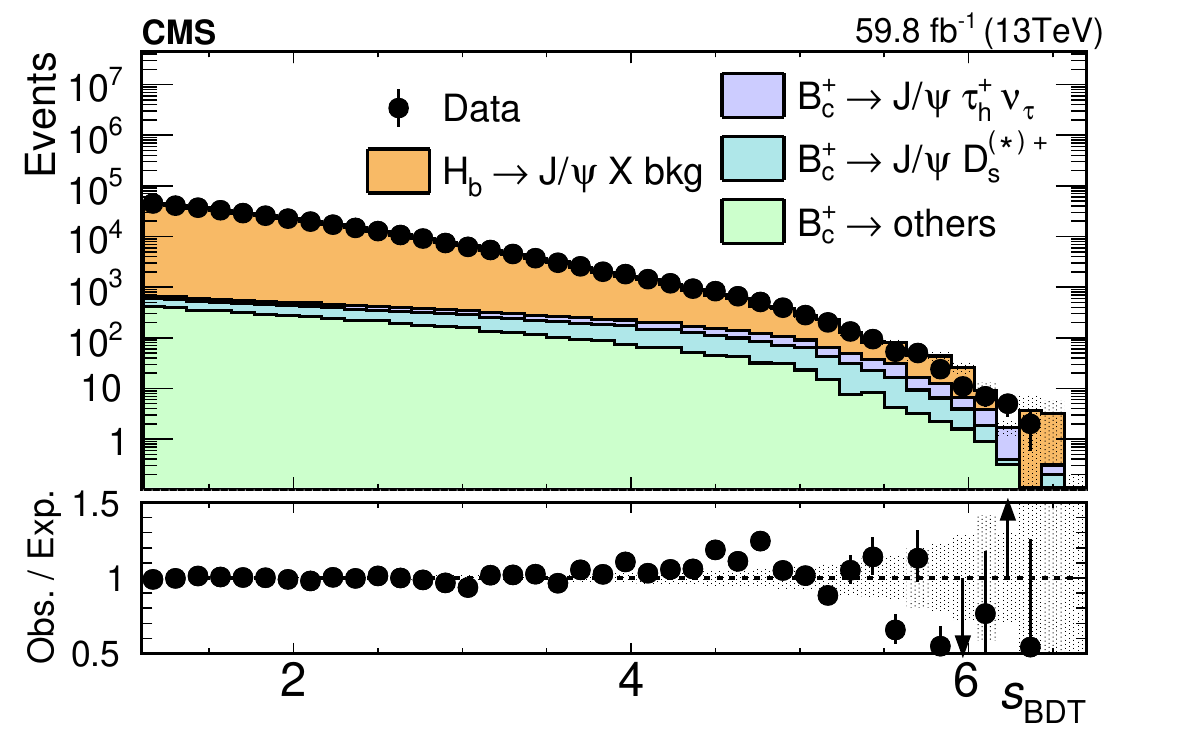}
\caption{The \bdt distribution. The bars on the data points represent the statistical uncertainties. The simulated \HbJpsiX contribution is normalized to match the yield in data after applying all event selection criteria except the BDT\@. The \bcjpsiX processes are normalized before the \rjpsi fit.
The lower panel displays the ratio between the data and the sum of expected signal and background contributions, with the statistical uncertainty in the sum of the simulated contributions shown by a gray band.
}
\label{fig:bdt}
\end{figure}

After these requirements, over 99.98\% of the events are inclusive \PBz and \PBp background events with a \JPsi candidate in the final state. 
We use a discriminator based on a boosted decision tree~(BDT) trained with the \textsc{XGBoost} algorithm~\cite{Chen:2016btl} to suppress the background. The \BcsignalTau simulated sample is used as signal for the training with the selected \PGt lepton candidate matched to the generator-level \tauHad. The simulated \HbJpsiX sample is used as the background process.
The BDT hyperparameters were optimized using a Bayesian technique and comparisons between data and simulation on control regions not used in the analysis were used to check the agreement. 
A training with background events from data was found to yield similar results.
The BDT uses 18 input variables related to the kinematic and topological properties of the \PB meson and \PGt lepton candidates, as well as global event-level observables. 
The \bdt distribution is depicted in Fig.~\ref{fig:bdt}, where expected simulated events are compared to 2018 data considering just the statistical uncertainties before the fit for \rjpsi. 

Using the \bdt variable, two regions are defined in each data-taking period: for the 2016 (2017--2018) data a signal-enriched region~(SR) with $\bdt > 3.5$ ($\bdt>4.2$) and a background-enriched sideband region~(SB) with $2<\bdt<3$ ($2.5<\bdt<3.5$). 
The SR region is optimized to maximize the signal significance, while the SB region is defined to include a sufficiently large number of events with \bdt values close enough to the SR to represent the background in that region, but far enough to minimize signal contamination.
In the SR, the signal and background efficiencies relative to the baseline selection are 10\% and $0.01\%$, respectively.
The 2016 and 2017--2018 \bdt regions differ because of a new pixel system installed after the 2016 data-taking period~\cite{Phase1Pixel}. 
For simulated signal events in the SR, 90\% of the \tauHad candidates are created from the three pions from the \PGt decay, while 10\% are created from two pions from the \PGt decay plus an unrelated track. Both are considered signal.

We exploit the dominant three-prong hadronic \PGt lepton decay \taualphapi~\cite{ALEPH:2005qgp} for the signal extraction, where \atorho. 
A one-dimensional distribution is built from the \rhoo vs\ \rhot invariant masses two-dimensional distribution, where \rhoo and \rhot are the two possible $\rhomass\!\to\!\pipi$ candidates from each \tauHad candidate ordered in \pt. We consider the mass range 0.2--1.5\GeV for \rhoo and \rhot, each split into four bins with widths of 0.225\GeV for the first three and 0.625\GeV for the last.
This is referred to as the unrolled $m(\rhoo)$ vs\ $m(\rhot)$ distribution.
Genuine \PGt lepton decays will tend to peak in either the \rhoo or \rhot invariant mass distribution. This is not the case for background events, which typically do not include \PGt lepton decays. The \rhoo and \rhot invariant masses are not included in the BDT and are found to be independent of \bdt.

Maximum likelihood fits are performed using the \textsc{Combine} tool~\cite{CMS:2024onh}. The data used by the fit are the same as used in the leptonic \PGt analysis~\cite{CMS:2023vgr} plus six unrolled $m(\rhoo)$ vs\ $m(\rhot)$ mass distributions (SB and SR for three data-taking periods). 
The signal strength parameter $r$ is proportional to the parameter of interest, \rjpsi and therefore only scales the \BcsignalTau yield. Another free parameter, \rBc, multiplies the initial normalization of all \PBpc MC events, including the \PBpc background as well as the \BcsignalMuon and \BcsignalTau contributions. The value of \rBc is primarily determined by the (large) number of \BcsignalMuon events. 
The fit model for the leptonic \PGt distributions is the same as in Ref.~\cite{CMS:2023vgr}. For the hadronic \PGt distributions,
the \PBpc MC sample is split into \BcsignalTau signal, \BcJpsiDs background, and other \PBpc backgrounds, with the unrolled $m(\rhoo)$ vs\ $m(\rhot)$ shapes for each contribution determined from simulation and the normalization set by $r$ and \rBc.

The expected yield of \HbJpsiX events in the $i$th bin of the unrolled $m(\rhoo)$ vs $m(\rhot)$ distribution in the SR is derived from the SB data after subtracting the expected number of \PBpc events:
\ifthenelse{\boolean{cms@external}}
{
\begin{multline}
  N_{\Hb} (\mathrm{SR},\,i) = \fexti [\, N_{\text{data}} (\mathrm{SB},\,i) - \\[-5pt]
  \rBc\,  N_\text{\PBpc,\,bkg} (\mathrm{SB},\,i)  -   \rBc\,  r\,  N_\text{\PBpc,\,sig} (\mathrm{SB},\,i) \,].
  \label{eq:hbjpsi_yield}
\end{multline}
}
{
\begin{equation}
  \label{eq:hbjpsi_yield}
  N_{\Hb} (\mathrm{SR},\,i) = \fexti \left[ N_{\text{data}} (\mathrm{SB},\,i) -  \rBc\,  N_\text{\PBpc,\,bkg} (\mathrm{SB},\,i)  -  \rBc\,  r\,  N_\text{\PBpc,\,sig} (\mathrm{SB},\,i) \right].
\end{equation}
}
The parameters $N_\text{\PBpc,\,bkg}$ and $N_\text{\PBpc,\,sig}$ denote the number of \PBpc background events (\BcJpsiDs and $\PBpc\!\to\!\text{others}$) and signal events, respectively. The \fexti factors, around 0.06--0.11, are calculated from simulated \HbJpsiX events to scale the \HbJpsiX background from the SB to the SR\@.  Before calculating \fexti, the simulated \HbJpsiX events are reweighted to account for differences in the data and MC \bdt distributions.  The reweighting is done by fitting the data to simulation \bdt ratio after subtracting the expected \PBpc contribution to the data, fitting this ratio in the region of $0<\bdt<3$ for 2016 and $0<\bdt<3.5$ for 2017--2018 samples, and using this fitted function to reweight all MC \HbJpsiX events in the SB and SR regions. 

To validate the background extrapolation method, two studies are performed using different control regions. First, we use the SB data to predict the backgrounds in the ``separation'' region between the SB and the SR, \ie, $\sbhighl < \bdt < \srlowl$ for 2016 and $\sbhigh < \bdt < \srlow$ for 2017--2018 samples. Second, we select events by inverting at least one requirement on the vertex probability and the displacement significance of the \PGt lepton in the SR and SB\@. The modified SB is used to predict the background in the modified SR\@. Agreement between the predicted and observed background is found in both cases.
In addition, the background predicted from the nominal SB is seen to agree with the simulated background within the statistical uncertainty in the simulation.

Systematic uncertainties for the normalization and shapes of the model are implemented as nuisance parameters in the fit.
Table~\ref{tab:sys} provides a summary of the systematic uncertainties, including information on the type of systematic uncertainty (affecting shape or normalization), which samples it affects, and its correlation across the different analysis channels. All systematic uncertainties are fully correlated between the SB and SR\@.

\begin{table*}[htb]
\centering
\topcaption{Systematic uncertainties considered in this analysis. The first, second, and third columns give the source, the type of effect (shape or normalization), and affected processes, respectively. A given uncertainty treated as fully correlated between analysis channels is marked with ``\checkmark''. Uncertainties that are irrelevant or negligible are marked with ``\NA''. \label{tab:sys}}
\cmsTable{\begin{scotch}{lclcccc} 
\multirow{2}{*}{Source}     & \multirow{2}{*}{Type} & \multirow{2}{*}{Affected process}       & \multicolumn{4}{c}{Channel}                                           \\ \cline{4-7}
                            &                       &                                       & $\PGt_\PGm$ 2018  & $\tauh$ 2018      & $\tauh$ 2017  & $\tauh$ 2016  \\ \hline
Form factor                 & shape                 & $\PBpc \!\to\! \JPsi \Pell\PGn_{\Pell}$   & \checkmark        & \checkmark        & \checkmark    & \checkmark    \\ 
\textsc{Tauola} model       & shape                 & \BcsignalTau                          &\NA        & \checkmark        & \checkmark    & \checkmark    \\
\HbJpsiX shape              & shape	                & \HbJpsiX (data)                       &\NA        & \checkmark        & \checkmark    & \checkmark    \\ 
Pileup weight               & shape                 & All MC                                & \checkmark        & \checkmark        & \checkmark    & \checkmark    \\ 
\PBpc lifetime              & shape                 & All \PBpc processes                      & \checkmark        & \checkmark        & \checkmark    & \checkmark    \\
$\PBpc\!\to\!\text{others}$     & shape                 & $\PBpc\!\to\!\text{others}$               &\NA        & \checkmark        & \checkmark    & \checkmark    \\
Finite MC size              & shape                 & All                                   & \checkmark        & \checkmark        & \checkmark    & \checkmark    \\ [\cmsTabSkip]
Triplet reconstruction efficiency         & normalization                 & \BcsignalTau &\NA & 6.9\% (\checkmark) &  6.9\% (\checkmark) &  6.9\% (\checkmark) \\
\BcJpsiDs normalization                  & normalization                 & \BcJpsiDs                             & 38\% (\checkmark) & 38\% (\checkmark) & 38\% (\checkmark) & 38\% (\checkmark)\\
$\PBpc\!\to\!\text{others}$ normalization     & normalization           & $\PBpc\!\to\!\text{others}$ & \NA & 50\% (\checkmark) & 50\% (\checkmark) & 50\% (\checkmark)\\
Trigger (\mumu)             & normalization                 & All MC                                & 10\% (\checkmark) & 10\%(\checkmark)$\oplus$ 5\% & 10\%& 10\%         \\
Trigger (track)             & normalization                 & All MC                                &\NA        & 10\%              & 10\%          & 10\%          \\
Trigger (\JPsi)             & normalization                 & All MC                                &\NA        & 10\%              & 10\%          & 10\%          \\
Muon ID efficiency                  & normalization                 & All MC                                & 4\%               & 4\%               & 4\%           & 4\%           \\
Muon reconstruction efficiency      & normalization                 & All MC                                & 4\% (\checkmark)  & 4\% (\checkmark)  & 4\%           & 4\%           \\ 
\HbJpsiX normalization              & normalization                 & \HbJpsiX (data)                       &\NA        & 30\%              & 30\%          & 30\%          \\ 
\PBpc MC normalization              & normalization                 & All \PBpc processes                   &\NA        & 5\%               & 30\%          & 30\%          \\ 
\end{scotch}}
\end{table*}

The \PBpc form factor uncertainty is evaluated with the \textsc{Hammer} framework~\cite{hammer} by considering ten variations that modify both the \BcsignalMuon and \BcsignalTau decays, with each variation corresponding to an eigenvector of the ten different BGL parameters~\cite{bgl:2019}.

While \taualphapi, \atorho dominates $\taualphapi\!\to\!\pipipi\PGnGt$ decays, the nonresonant $\aresonance\!\to\!\pipipi$ decay has also been observed, and the relative rates are unknown~\cite{ParticleDataGroup:2022pth}. To account for this, generator-level events are simulated with an alternative \textsc{Tauola} setting (changing ``TauolaCurrentOption'' from 1 (BaBar-tuned) to 0 (CLEO-tuned)), and the resulting signal template is built by reweighting events based on the generator-level unrolled $m(\rhoo)$ vs\ $m(\rhot)$ distribution.

The \HbJpsiX background is estimated from data in the SB region.
Possible shape differences in the \rhoz mass distributions between SB and SR are corrected via the bin-by-bin extrapolation factors \fexti, which are obtained from simulation. These factors are varied within their uncertainties and the resulting templates are included as a systematic uncertainty. They are treated as uncorrelated between all bins of the unrolled $m(\rhoo)$ vs\ $m(\rhot)$ distribution, as they are dominated by the statistical uncertainties of the simulated events. The uncertainties are typically in the 5--20\% range.

The variation of the \PBpc lifetime within its measured uncertainty is applied to all \PBpc decays.

The pileup modeling uncertainty is assessed for all simulated processes by varying the total inelastic cross section used to generate the pileup distributions~\cite{Sirunyan:2018nqx}.

As previously described, the unknown contributions to $\PBpc\!\to\!\text{others}$ are accounted for by scaling up the generated MC events. Uncertainties on the normalization (amounting to half of the scaling) and shape are assessed.

The statistical uncertainties of the simulated events are included independently in each bin. This uncertainty is also propagated to the \HbJpsiX background estimate from the subtraction in Eq.~\eqref{eq:hbjpsi_yield}. 

The hadronic \PGt decay contains three pions compared to the normalization channel of one muon. Therefore, discrepancies in the track reconstruction efficiency between data and simulation will affect the result.
A 2.3\% uncertainty~\cite{tracking} is assumed per track, yielding a total normalization uncertainty of 6.9\% affecting the \BcsignalTau process.
A check is performed by measuring the yields of the \bplustojpsikplus and \bplustopsiprimekplus processes, as the only difference between the two processes is the track multiplicity.
The ratios of the yields in data and simulation differ by 4.4\%, corresponding to an uncertainty of 2.2\% per track, compatible with the assumed 2.3\% uncertainty per track.

The normalization uncertainty for the \BcJpsiDs background originates from experimentally measured branching fractions as in the leptonic \PGt analysis~\cite{CMS:2023vgr}.

Uncertainties in the efficiencies of the muon reconstruction (4\%), ID (4\%), and trigger selection (10\%) are considered. These uncertainties are considered uncorrelated across data-taking years. 
For 2018, the muon trigger uncertainties are partly correlated between this analysis and the leptonic \PGt analysis because the event selection is similar, with an additional uncorrelated 5\% uncertainty owing to differences in online muon selection. The muon reconstruction uncertainty is also correlated, whereas the muon ID uncertainty is uncorrelated as different criteria are used. 
Trigger efficiency uncertainties related to the track reconstruction (10\%) and the \JPsi candidate vertex displacement requirement (10\%) are also considered.

Given the uncertainty in the check of the backgrounds predicted from the SBs in the ``separation'' regions and those with the inverted \PGt lepton selection, a 30\% uncertainty is assigned to the \HbJpsiX background normalization, uncorrelated across years.

The normalization of the \PBpc MC samples for 2016 and 2017 relative to 2018 is obtained from the ratio of the yields of \BcJpsiPiPiPi events in data and MC in each year. These relative correction factors for 2016 and 2017 are also calculated using the \BcJpsiPi invariant mass peak, and the difference with respect to the ones from \BcJpsiPiPiPi is used to determine a \PBpc MC normalization systematic uncertainty of 30\% for each of 2016 and 2017 periods, uncorrelated by year.

The leading systematic uncertainties for the hadronic \Pgt analysis are from the \HbJpsiX background estimation, obtained from data, and its extrapolation from the SB to the SR, which is based on simulation.

Several fits have been performed using different subsets of the data. As the hadronic \PGt analysis relies on the leptonic \PGt data to determine the \BcsignalMuon yield, we cannot obtain a completely independent hadronic \PGt value for \rjpsi. However, we can fit the hadronic \PGt and leptonic \PGt data with separate \rjpsi parameters for each. In this case, the two results share a common denominator (\BcsignalMuon yield), but the \BcsignalTau yields are independent. Incorporating just the 2018 data to match the leptonic \PGt analysis, this fit returns $\rjpsihadeighteen=0.74_{-0.53}^{+0.57}$. Including the hadronic \PGt decay data recorded in 2016 and 2017, a value of $\rjpsihad=1.04_{-0.44}^{+0.50}$ is obtained. We verified that this result is stable against inflating the prefit uncertainty in the \bcjpsiX normalization.
Given the uncertainties, this result is compatible with that of the leptonic \PGt analysis, $0.17 \pm 0.33$. We compared the likelihoods of the fits to the data of this analysis and to the data of the leptonic \PGt analysis when assuming one common parameter $r$ or two independent parameters $r_{\text{lep}}$ and $r_{\text{had}}$. The likelihood ratio, expressed as ${-}2\ln \mathcal{L} (\text{data}|r) / \mathcal{L} (\text{data}|r_{\text{lep}},\, r_{\text{had}})$, is 2.06, corresponding to a $p$-value \cite{Fisher1992} of 0.15, indicating that the two measurements are compatible. 

\begin{figure}[htb!]
\centering
\includegraphics[width=0.49\textwidth]{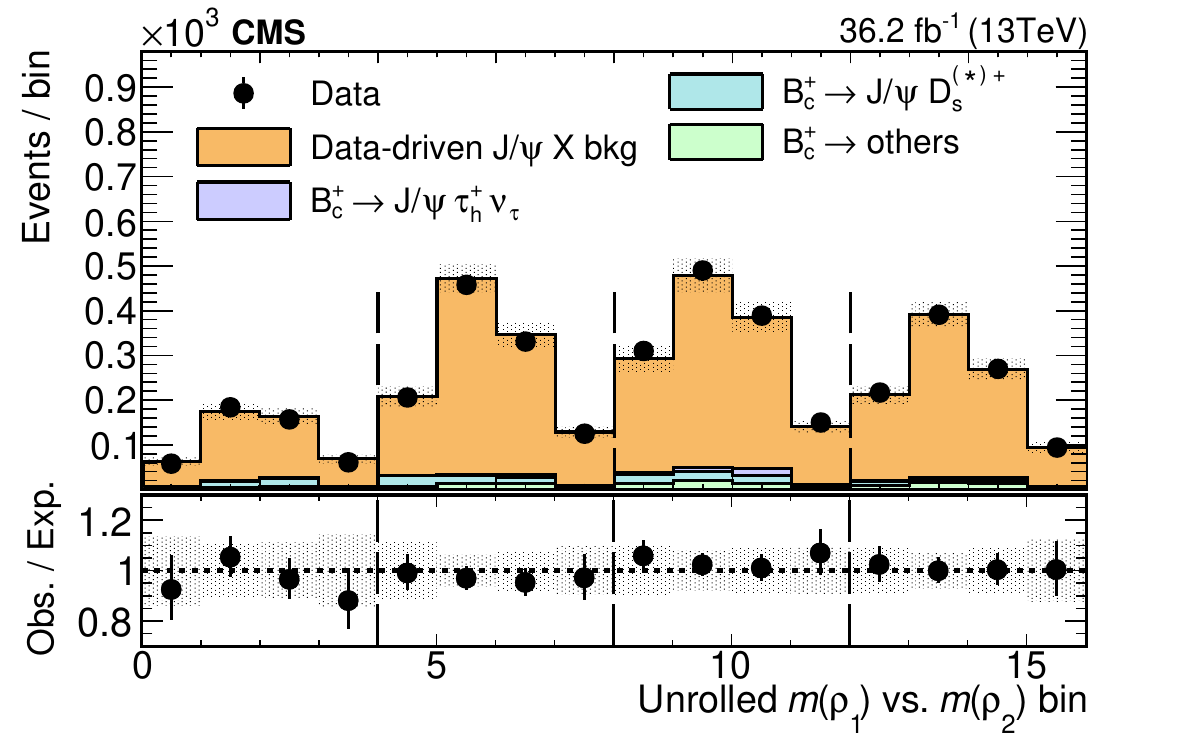}
\includegraphics[width=0.49\textwidth]{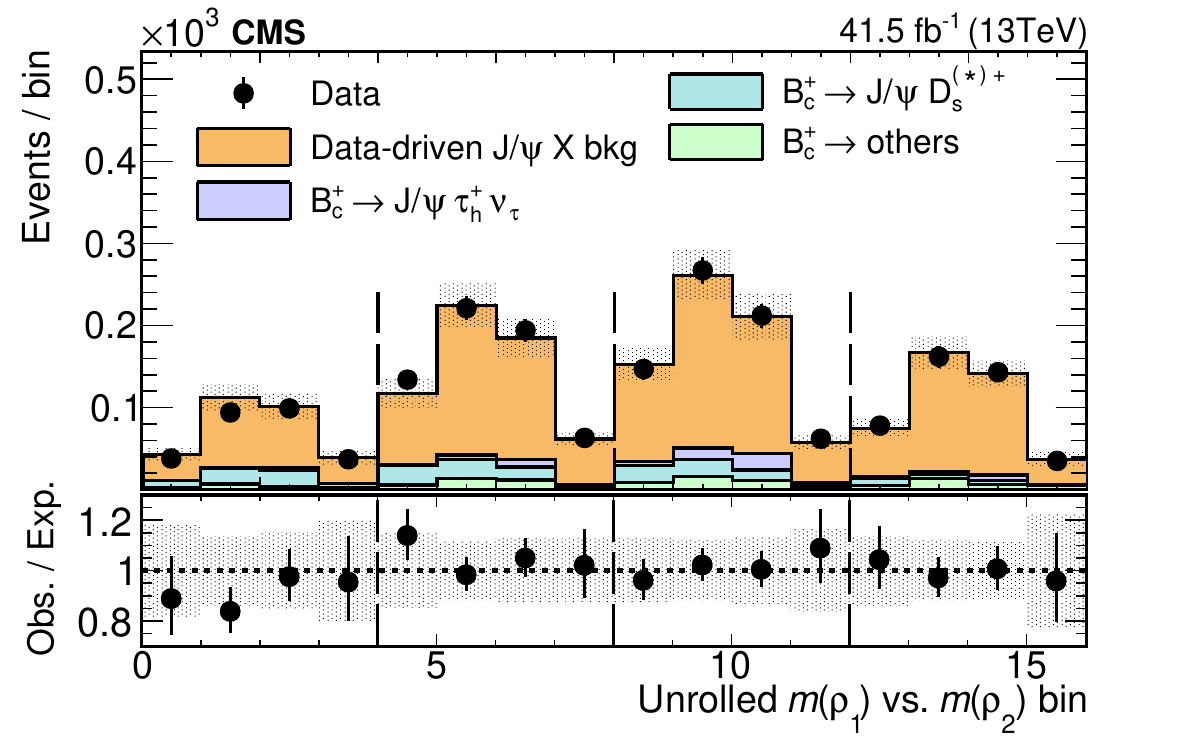}
\includegraphics[width=0.49\textwidth]{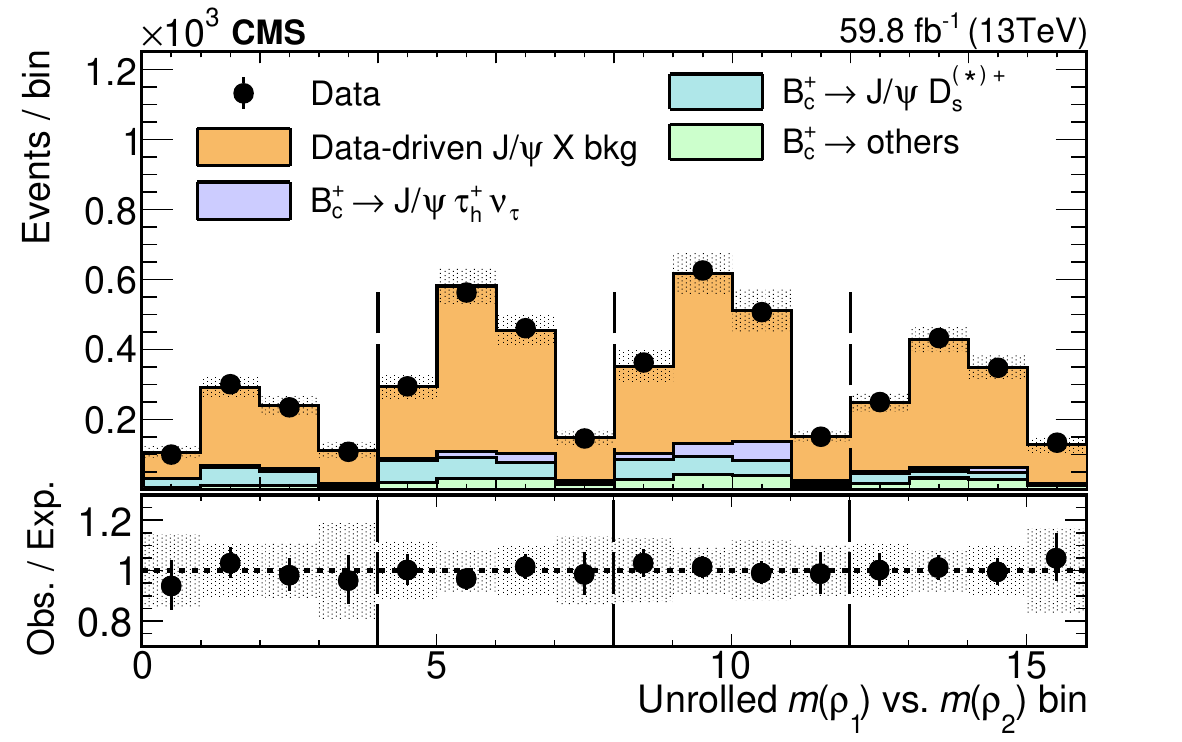}
\caption{
The SR unrolled $m(\rhoo)$ vs\ $m(\rhot)$ distributions for the 2016 (\cmsLeft), 2017 (\cmsMiddle), and 2018 (\cmsRight) data-taking periods. The dashed vertical lines indicate bins in $m(\rhoo)$. Histograms show the postfit expected yields of the \HbJpsiX background estimated from data and simulated \BcsignalTau, \BcJpsiDs, and $\PBpc\!\to\!\text{others}$ processes. The ratio between the data and the sum of postfit signal and background is shown in the lower panel, with a gray band showing the total expected event yield uncertainty.
\label{fig:unrolledmass}}
\end{figure}

\begin{figure}[htb!]
\centering
\includegraphics[width=\cmsFigSize]{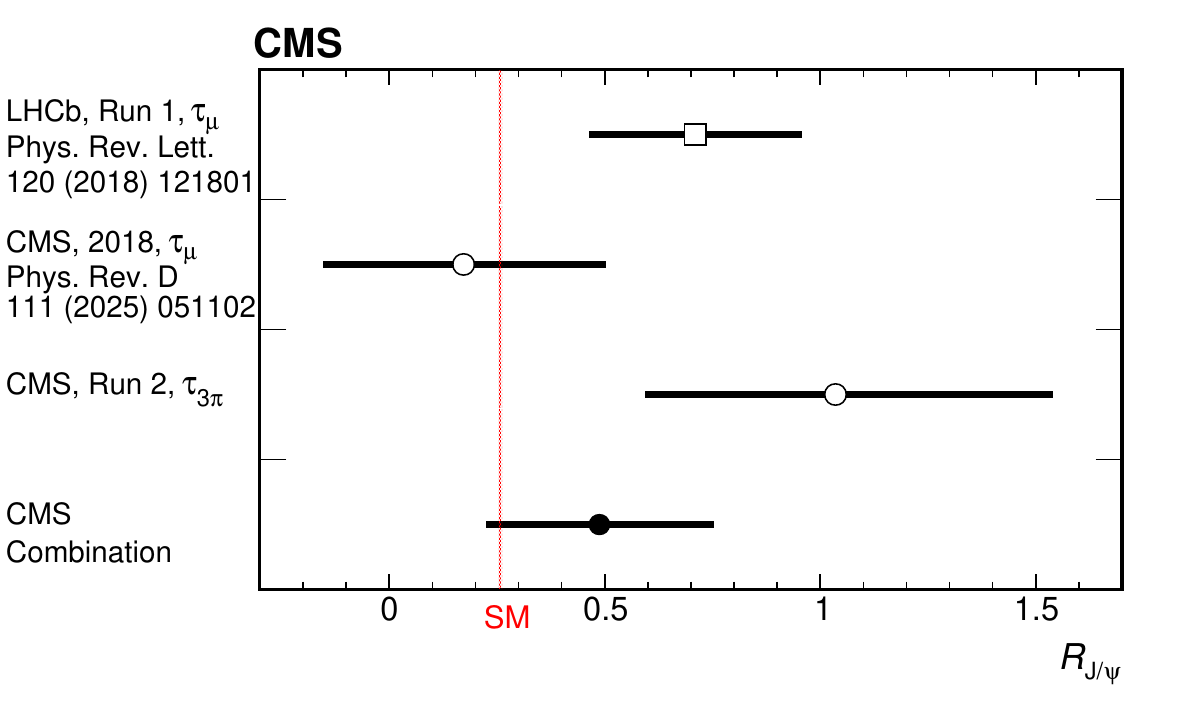}
\caption{Summary of \rjpsi measurements from LHCb leptonic \PGt decays~\cite{LHCb:2017vlu}, CMS leptonic \PGt decays~\cite{CMS:2023vgr}, CMS hadronic \PGt decays, and the CMS combination. The total uncertainties in the measured values are indicated by the horizontal lines. The vertical red line is the SM prediction.
\label{fig:rjpsiresult}}
\end{figure}

Finally, we perform a simultaneous fit in all analysis regions of this and of the leptonic \PGt analysis to obtain \rjpsi = \result,
where the last uncertainty corresponds to the \PBpc form factors and the systematic uncertainty includes the other systematic uncertainties.
Figure~\ref{fig:unrolledmass} shows the SR unrolled $m(\rhoo)$ vs\ $m(\rhot)$ distributions of the data and of the final fit for each of the three data-taking years.
The results agrees with the SM prediction of \theoryresult, as shown in Fig.~\ref{fig:rjpsiresult}, where a summary of the CMS measurements in each \PGt decay channel, their combination, the SM prediction, and the LHCb result~\cite{LHCb:2017vlu} are presented. 

In summary, the branching fraction ratio \RJpsi has been measured using \jpsimumu decays for both the numerator and denominator, and the three-prong hadronic \PGt decay 
for the numerator.
Proton-proton collision data at $\sqrt{s} = 13\TeV$ collected by the CMS experiment are used, corresponding to an integrated luminosity of 138\fbinv for the numerator and 59.7\fbinv for the denominator. This is the first contribution to lepton flavor universality tests in \PBpc meson decays with hadronic \Pgt decays. The ratio is measured to be $\rjpsihad=1.04_{-0.44}^{+0.50}$. Combined with our previous measurement using leptonic \PGt decays~\cite{CMS:2023vgr}, the result is $\rjpsi=\resultc$, consistent with the standard model prediction of \theoryresult.

\begin{acknowledgments}
We congratulate our colleagues in the CERN accelerator departments for the excellent performance of the LHC and thank the technical and administrative staffs at CERN and at other CMS institutes for their contributions to the success of the CMS effort. In addition, we gratefully acknowledge the computing centers and personnel of the Worldwide LHC Computing Grid and other centers for delivering so effectively the computing infrastructure essential to our analyses. Finally, we acknowledge the enduring support for the construction and operation of the LHC, the CMS detector, and the supporting computing infrastructure provided by the following funding agencies: SC (Armenia), BMBWF and FWF (Austria); FNRS and FWO (Belgium); CNPq, CAPES, FAPERJ, FAPERGS, and FAPESP (Brazil); MES and BNSF (Bulgaria); CERN; CAS, MoST, and NSFC (China); MINCIENCIAS (Colombia); MSES and CSF (Croatia); RIF (Cyprus); SENESCYT (Ecuador); ERC PRG, TARISTU24-TK10 and MoER TK202 (Estonia); Academy of Finland, MEC, and HIP (Finland); CEA and CNRS/IN2P3 (France); SRNSF (Georgia); BMFTR, DFG, and HGF (Germany); GSRI (Greece); NKFIH (Hungary); DAE and DST (India); IPM (Iran); SFI (Ireland); INFN (Italy); MSIT and NRF (Republic of Korea); MES (Latvia); LMTLT (Lithuania); MOE and UM (Malaysia); BUAP, CINVESTAV, CONACYT, LNS, SEP, and UASLP-FAI (Mexico); MOS (Montenegro); MBIE (New Zealand); PAEC (Pakistan); MES, NSC, and NAWA (Poland); FCT (Portugal); MESTD (Serbia); MICIU/AEI and PCTI (Spain); MOSTR (Sri Lanka); Swiss Funding Agencies (Switzerland); MST (Taipei); MHESI (Thailand); TUBITAK and TENMAK (T\"{u}rkiye); NASU (Ukraine); STFC (United Kingdom); DOE and NSF (USA).

\hyphenation{Rachada-pisek} Individuals have received support from the Marie-Curie program and the European Research Council and Horizon 2020 Grant, contract Nos.\ 675440, 724704, 752730, 758316, 765710, 824093, 101115353, 101002207, 101001205, and COST Action CA16108 (European Union); the Leventis Foundation; the Alfred P.\ Sloan Foundation; the Alexander von Humboldt Foundation; the Science Committee, project no. 22rl-037 (Armenia); the Fonds pour la Formation \`a la Recherche dans l'Industrie et dans l'Agriculture (FRIA) and Fonds voor Wetenschappelijk Onderzoek contract No. 1228724N (Belgium); the Beijing Municipal Science \& Technology Commission, No. Z191100007219010, the Fundamental Research Funds for the Central Universities, the Ministry of Science and Technology of China under Grant No. 2023YFA1605804, the Natural Science Foundation of China under Grant No. 12061141002, 12535004, and USTC Research Funds of the Double First-Class Initiative No.\ YD2030002017 (China); the Ministry of Education, Youth and Sports (MEYS) of the Czech Republic; the Shota Rustaveli National Science Foundation, grant FR-22-985 (Georgia); the Deutsche Forschungsgemeinschaft (DFG), among others, under Germany's Excellence Strategy -- EXC 2121 ``Quantum Universe" -- 390833306, and under project number 400140256 - GRK2497; the Hellenic Foundation for Research and Innovation (HFRI), Project Number 2288 (Greece); the Hungarian Academy of Sciences, the New National Excellence Program - \'UNKP, the NKFIH research grants K 131991, K 133046, K 138136, K 143460, K 143477, K 146913, K 146914, K 147048, 2020-2.2.1-ED-2021-00181, TKP2021-NKTA-64, and 2021-4.1.2-NEMZ\_KI-2024-00036 (Hungary); the Council of Science and Industrial Research, India; ICSC -- National Research Center for High Performance Computing, Big Data and Quantum Computing, FAIR -- Future Artificial Intelligence Research, and CUP I53D23001070006 (Mission 4 Component 1), funded by the NextGenerationEU program (Italy); the Latvian Council of Science; the Ministry of Education and Science, project no. 2022/WK/14, and the National Science Center, contracts Opus 2021/41/B/ST2/01369, 2021/43/B/ST2/01552, 2023/49/B/ST2/03273, and the NAWA contract BPN/PPO/2021/1/00011 (Poland); the Funda\c{c}\~ao para a Ci\^encia e a Tecnologia, grant CEECIND/01334/2018 (Portugal); the National Priorities Research Program by Qatar National Research Fund; MICIU/AEI/10.13039/501100011033, ERDF/EU, "European Union NextGenerationEU/PRTR", and Programa Severo Ochoa del Principado de Asturias (Spain); the Chulalongkorn Academic into Its 2nd Century Project Advancement Project, the National Science, Research and Innovation Fund program IND\_FF\_68\_369\_2300\_097, and the Program Management Unit for Human Resources \& Institutional Development, Research and Innovation, grant B39G680009 (Thailand); the Kavli Foundation; the Nvidia Corporation; the SuperMicro Corporation; the Welch Foundation, contract C-1845; and the Weston Havens Foundation (USA).
\end{acknowledgments}

\bibliography{auto_generated} 

\providecommand{\href}[2]{#2}\begingroup\raggedright\begin{thebibliography}{10}%
\makeatletter
\providecommand{\hrefCMSnoop }[0]{\@secondoftwo}%
\makeatother
\providecommand{\doi}{\texttt{doi:}\begingroup \urlstyle{tt}\Url}

\bibitem{ATLAS:2012yve}
\hrefCMSnoop {}{{ATLAS Collaboration}, ``Observation of a new particle in the
  search for the standard model {Higgs} boson with the {ATLAS} detector at the
  {LHC}'',} \textit{ Phys. Lett. B} \textbf{ 716} (2012) 1,
  \href{http://dx.doi.org/10.1016/j.physletb.2012.08.020}{\doi{10.1016/j.physletb.2012.08.020}},
  \href{http://www.arXiv.org/abs/1207.7214}{\texttt{arXiv:1207.7214}}.

\bibitem{CMS:2012qbp}
\hrefCMSnoop {}{{CMS Collaboration}, ``Observation of a new boson at a mass of
  125 {GeV} with the {CMS} experiment at the {LHC}'',} \textit{ Phys. Lett. B}
  \textbf{ 716} (2012) 30,
  \href{http://dx.doi.org/10.1016/j.physletb.2012.08.021}{\doi{10.1016/j.physletb.2012.08.021}},
  \href{http://www.arXiv.org/abs/1207.7235}{\texttt{arXiv:1207.7235}}.

\bibitem{CMS:2013btf}
\hrefCMSnoop {}{{CMS Collaboration}, ``Observation of a new boson with mass
  near 125 {GeV} in {$\Pp\Pp$} collisions at $\sqrt{s}$ = 7 and 8\,{TeV}'',}
  \textit{ JHEP} \textbf{ 06} (2013) 081,
  \href{http://dx.doi.org/10.1007/JHEP06(2013)081}{\doi{10.1007/JHEP06(2013)081}},
  \href{http://www.arXiv.org/abs/1303.4571}{\texttt{arXiv:1303.4571}}.

\bibitem{PhysRevLett.10.531}
\hrefCMSnoop {}{N.~Cabibbo, ``Unitary symmetry and leptonic decays'',} \textit{
  Phys. Rev. Lett.} \textbf{ 10} (1963) 531,
  \href{http://dx.doi.org/10.1103/PhysRevLett.10.531}{\doi{10.1103/PhysRevLett.10.531}}.

\bibitem{Kobayashi:1973fv}
\hrefCMSnoop {}{M.~Kobayashi and T.~Maskawa, ``{CP} violation in the
  renormalizable theory of weak interaction'',} \textit{ Prog. Theor. Phys.}
  \textbf{ 49} (1973) 652,
  \href{http://dx.doi.org/10.1143/PTP.49.652}{\doi{10.1143/PTP.49.652}}.

\bibitem{CMS:2018fdh}
\hrefCMSnoop {}{{CMS Collaboration}, ``Evidence for associated production of a
  {Higgs} boson with a top quark pair in final states with electrons, muons,
  and hadronically decaying $\tau$ leptons at $\sqrt{s} =$ 13\,{TeV}'',}
  \textit{ JHEP} \textbf{ 08} (2018) 066,
  \href{http://dx.doi.org/10.1007/JHEP08(2018)066}{\doi{10.1007/JHEP08(2018)066}},
  \href{http://www.arXiv.org/abs/1803.05485}{\texttt{arXiv:1803.05485}}.

\bibitem{CMS:2018nsn}
\hrefCMSnoop {}{{CMS Collaboration}, ``Observation of {Higgs} boson decay to
  bottom quarks'',} \textit{ Phys. Rev. Lett.} \textbf{ 121} (2018) 121801,
  \href{http://dx.doi.org/10.1103/PhysRevLett.121.121801}{\doi{10.1103/PhysRevLett.121.121801}},
  \href{http://www.arXiv.org/abs/1808.08242}{\texttt{arXiv:1808.08242}}.

\bibitem{LHCb:2017vlu}
\hrefCMSnoop {}{{LHCb Collaboration}, ``Measurement of the ratio of branching
  fractions
  {$\mathcal{B}(\PBpc\to\JPsi\PGtp\PGnGt)/\mathcal{B}(\PBpc\to\JPsi\PGmp\PGnGm)$}'',}
  \textit{ Phys. Rev. Lett.} \textbf{ 120} (2018) 121801,
  \href{http://dx.doi.org/10.1103/PhysRevLett.120.121801}{\doi{10.1103/PhysRevLett.120.121801}},
  \href{http://www.arXiv.org/abs/1711.05623}{\texttt{arXiv:1711.05623}}.

\bibitem{Harrison:2020nrv}
\hrefCMSnoop {}{{LATTICE-HPQCD} Collaboration, ``{$\mathcal{R}(\JPsi)$} and
  {$\PBpc\to\JPsi \ell^-\overline{\PGn}_\ell$} lepton flavor universality
  violating observables from lattice {QCD}'',} \textit{ Phys. Rev. Lett.}
  \textbf{ 125} (2020) 222003,
  \href{http://dx.doi.org/10.1103/PhysRevLett.125.222003}{\doi{10.1103/PhysRevLett.125.222003}},
  \href{http://www.arXiv.org/abs/2007.06956}{\texttt{arXiv:2007.06956}}.

\bibitem{Anisimov:1998uk}
\hrefCMSnoop {}{A.~Y. Anisimov, I.~M. Narodetsky, C.~Semay, and
  B.~Silvestre-Brac, ``The {\PBpc} meson lifetime in the light front
  constituent quark model'',} \textit{ Phys. Lett. B} \textbf{ 452} (1999) 129,
  \href{http://dx.doi.org/10.1016/S0370-2693(99)00273-7}{\doi{10.1016/S0370-2693(99)00273-7}},
  \href{http://www.arXiv.org/abs/hep-ph/9812514}{\texttt{arXiv:hep-ph/9812514}}.

\bibitem{Ivanov:2006ni}
\hrefCMSnoop {}{M.~A. Ivanov, J.~G. Korner, and P.~Santorelli, ``Exclusive
  semileptonic and nonleptonic decays of the {\PBpc} meson'',} \textit{ Phys.
  Rev. D} \textbf{ 73} (2006) 054024,
  \href{http://dx.doi.org/10.1103/PhysRevD.73.054024}{\doi{10.1103/PhysRevD.73.054024}},
  \href{http://www.arXiv.org/abs/hep-ph/0602050}{\texttt{arXiv:hep-ph/0602050}}.

\bibitem{Hernandez:2006gt}
\hrefCMSnoop {}{E.~Hernandez, J.~Nieves, and J.~M. Verde-Velasco, ``Study of
  exclusive semileptonic and non-leptonic decays of {\PBmc} in a
  nonrelativistic quark model'',} \textit{ Phys. Rev. D} \textbf{ 74} (2006)
  074008,
  \href{http://dx.doi.org/10.1103/PhysRevD.74.074008}{\doi{10.1103/PhysRevD.74.074008}},
  \href{http://www.arXiv.org/abs/hep-ph/0607150}{\texttt{arXiv:hep-ph/0607150}}.

\bibitem{Azizi:2019aaf}
\hrefCMSnoop {}{K.~Azizi, Y.~Sarac, and H.~Sundu, ``Lepton flavor universality
  violation in semileptonic tree level weak transitions'',} \textit{ Phys. Rev.
  D} \textbf{ 99} (2019) 113004,
  \href{http://dx.doi.org/10.1103/PhysRevD.99.113004}{\doi{10.1103/PhysRevD.99.113004}},
  \href{http://www.arXiv.org/abs/1904.08267}{\texttt{arXiv:1904.08267}}.

\bibitem{CMS:2023vgr}
\hrefCMSnoop {}{{CMS Collaboration}, ``Test of lepton flavor universality in
  semileptonic {\PBpc} meson decays in proton-proton collisions at $\sqrt{s}$ =
  13\,{TeV}'',} \textit{ Phys. Rev. D} \textbf{ 111} (2025) L051102,
  \href{http://dx.doi.org/10.1103/PhysRevD.111.L051102}{\doi{10.1103/PhysRevD.111.L051102}},
  \href{http://www.arXiv.org/abs/2408.00678}{\texttt{arXiv:2408.00678}}.

\bibitem{CMS-PAS-LUM-18-002}
\href {https://cds.cern.ch/record/2676164/}{{CMS Collaboration}, ``{CMS}
  luminosity measurement for the 2018 data-taking period at $\sqrt{s}$ =
  13\,{TeV}'',} CMS Physics Analysis Summary CMS-PAS-LUM-18-002, 2019.

\bibitem{hepdata}
\hrefCMSnoop {}{}{HEPD}ata record for this analysis, 2025.
\newblock
  \href{http://dx.doi.org/10.17182/hepdata.166008}{\doi{10.17182/hepdata.166008}}.

\bibitem{CMS-PAS-LUM-17-004}
\href {https://cds.cern.ch/record/2621960/}{{CMS Collaboration}, ``{CMS}
  luminosity measurement for the 2017 data-taking period at $\sqrt{s}$ =
  13\,{TeV}'',} CMS Physics Analysis Summary CMS-PAS-LUM-17-004, 2018.

\bibitem{CMS:2021xjt}
\hrefCMSnoop {}{{CMS Collaboration}, ``Precision luminosity measurement in
  proton-proton collisions at $\sqrt{s} =$ 13\,{TeV} in 2015 and 2016 at
  {CMS}'',} \textit{ Eur. Phys. J. C} \textbf{ 81} (2021) 800,
  \href{http://dx.doi.org/10.1140/epjc/s10052-021-09538-2}{\doi{10.1140/epjc/s10052-021-09538-2}},
  \href{http://www.arXiv.org/abs/2104.01927}{\texttt{arXiv:2104.01927}}.

\bibitem{CMS:2008xjf}
\hrefCMSnoop {}{{CMS Collaboration}, ``The {CMS} experiment at the {CERN
  LHC}'',} \textit{ JINST} \textbf{ 3} (2008) S08004,
  \href{http://dx.doi.org/10.1088/1748-0221/3/08/S08004}{\doi{10.1088/1748-0221/3/08/S08004}}.

\bibitem{CMS:2023gfb}
\hrefCMSnoop {}{{CMS Collaboration}, ``Development of the {CMS} detector for
  the {CERN LHC Run 3}'',} \textit{ JINST} \textbf{ 19} (2024) P05064,
  \href{http://dx.doi.org/10.1088/1748-0221/19/05/P05064}{\doi{10.1088/1748-0221/19/05/P05064}}.

\bibitem{CMS:2020cmk}
\hrefCMSnoop {}{{CMS Collaboration}, ``Performance of the {CMS Level-1} trigger
  in proton-proton collisions at $\sqrt{s} = 13$\,{TeV}'',} \textit{ JINST}
  \textbf{ 15} (2020) P10017,
  \href{http://dx.doi.org/10.1088/1748-0221/15/10/P10017}{\doi{10.1088/1748-0221/15/10/P10017}},
  \href{http://www.arXiv.org/abs/2006.10165}{\texttt{arXiv:2006.10165}}.

\bibitem{CMS:2016ngn}
\hrefCMSnoop {}{{CMS Collaboration}, ``The {CMS} trigger system'',} \textit{
  JINST} \textbf{ 12} (2017) P01020,
  \href{http://dx.doi.org/10.1088/1748-0221/12/01/P01020}{\doi{10.1088/1748-0221/12/01/P01020}},
\href{http://www.arXiv.org/abs/1609.02366}{\texttt{arXiv:1609.02366}}.

\bibitem{CMS:2024aqx}
\hrefCMSnoop {}{{CMS Collaboration}, ``Performance of the {CMS} high-level
  trigger during {LHC} {Run} 2'',} \textit{ JINST} \textbf{ 19} (2024) P11021,
  \href{http://dx.doi.org/10.1088/1748-0221/19/11/P11021}{\doi{10.1088/1748-0221/19/11/P11021}},
  \href{http://www.arXiv.org/abs/2410.17038}{\texttt{arXiv:2410.17038}}.

\bibitem{CMS:2020uim}
\hrefCMSnoop {}{{CMS Collaboration}, ``Electron and photon reconstruction and
  identification with the {CMS} experiment at the {CERN} {LHC}'',} \textit{
  JINST} \textbf{ 16} (2021) P05014,
  \href{http://dx.doi.org/10.1088/1748-0221/16/05/P05014}{\doi{10.1088/1748-0221/16/05/P05014}},
  \href{http://www.arXiv.org/abs/2012.06888}{\texttt{arXiv:2012.06888}}.

\bibitem{CMS:2018rym}
\hrefCMSnoop {}{{CMS Collaboration}, ``Performance of the {CMS} muon detector
  and muon reconstruction with proton-proton collisions at $\sqrt{s}=$
  13\,{TeV}'',} \textit{ JINST} \textbf{ 13} (2018) P06015,
  \href{http://dx.doi.org/10.1088/1748-0221/13/06/P06015}{\doi{10.1088/1748-0221/13/06/P06015}},
  \href{http://www.arXiv.org/abs/1804.04528}{\texttt{arXiv:1804.04528}}.

\bibitem{CMS:2014pgm}
\hrefCMSnoop {}{{CMS Collaboration}, ``Description and performance of track and
  primary-vertex reconstruction with the {CMS} tracker'',} \textit{ JINST}
  \textbf{ 9} (2014) P10009,
  \href{http://dx.doi.org/10.1088/1748-0221/9/10/P10009}{\doi{10.1088/1748-0221/9/10/P10009}},
  \href{http://www.arXiv.org/abs/1405.6569}{\texttt{arXiv:1405.6569}}.

\bibitem{CMS:2017yfk}
\hrefCMSnoop {}{{CMS Collaboration}, ``Particle-flow reconstruction and global
  event description with the {CMS} detector'',} \textit{ JINST} \textbf{ 12}
  (2017) P10003,
  \href{http://dx.doi.org/10.1088/1748-0221/12/10/P10003}{\doi{10.1088/1748-0221/12/10/P10003}},
\href{http://www.arXiv.org/abs/1706.04965}{\texttt{arXiv:1706.04965}}.

\bibitem{CMS:2018jrd}
\hrefCMSnoop {}{{CMS Collaboration}, ``Performance of reconstruction and
  identification of $\tau$ leptons decaying to hadrons and $\nu_\tau$ in pp
  collisions at $\sqrt{s}=$ 13\,{TeV}'',} \textit{ JINST} \textbf{ 13} (2018)
  P10005,
  \href{http://dx.doi.org/10.1088/1748-0221/13/10/P10005}{\doi{10.1088/1748-0221/13/10/P10005}},
  \href{http://www.arXiv.org/abs/1809.02816}{\texttt{arXiv:1809.02816}}.

\bibitem{Bcveg1}
\hrefCMSnoop {}{C.~Chang, C.~Driouchi, P.~Eerola, and X.~Wu, ``{BCVEGPY}: an
  event generator for hadronic production of the {$\PBpc$} meson'',} \textit{
  Comput. Phys. Commun.} \textbf{ 159} (2004) 33,
  \href{http://dx.doi.org/10.1016/j.cpc.2004.02.005}{\doi{10.1016/j.cpc.2004.02.005}},
\href{http://www.arXiv.org/abs/hep-ph/0309120}{\texttt{arXiv:hep-ph/0309120}}.

\bibitem{Bcveg}
\hrefCMSnoop {}{C.-H. Chang, X.-Y. Wang, and X.-G. Wu, ``{BCVEGPY2.2:} a newly
  upgraded version for hadronic production of the meson {$\rm B_c$} and its
  excited states'',} \textit{ Comput. Phys. Commun.} \textbf{ 197} (2015) 335,
  \href{http://dx.doi.org/10.1016/j.cpc.2015.07.015}{\doi{10.1016/j.cpc.2015.07.015}},
\href{http://www.arXiv.org/abs/1507.05176}{\texttt{arXiv:1507.05176}}.

\bibitem{Ryd:2005zz}
\hrefCMSnoop {}{D.~J. Lange, ``The {EvtGen} particle decay simulation
  package'',} \textit{ Nucl. Instrum. Meth. A} \textbf{ 462} (2001) 152,
  \href{http://dx.doi.org/10.1016/S0168-9002(01)00089-4}{\doi{10.1016/S0168-9002(01)00089-4}}.

\bibitem{berezhnoy2011bcnpimoduleanalysisbc}
\hrefCMSnoop {}{A.~V. Berezhnoy, A.~K. Likhoded, and A.~V. Luchinsky,
  ``{BC\_NPI} module for the analysis of {$\PB_\mathrm{c} \rightarrow \JPsi +
  \mathrm{n}\pi$} and {$\PB_\mathrm{c} \rightarrow \PB_\mathrm{s} +
  \mathrm{n}\pi$} decays within the {EvtGen} package'',} 2011.
  \href{http://www.arXiv.org/abs/1104.0808}{\texttt{arXiv:1104.0808}}.

\bibitem{Sjostrand:2014zea}
T.~Sj{\"o}strand\hrefCMSnoop {}{ { et~al.}, ``An introduction to {PYTHIA}
  8.2'',} \textit{ Comput. Phys. Commun.} \textbf{ 191} (2015) 159,
  \href{http://dx.doi.org/10.1016/j.cpc.2015.01.024}{\doi{10.1016/j.cpc.2015.01.024}},
  \href{http://www.arXiv.org/abs/1410.3012}{\texttt{arXiv:1410.3012}}.

\bibitem{CMS:2019csb}
\hrefCMSnoop {}{{CMS Collaboration}, ``Extraction and validation of a new set
  of {CMS} {PYTHIA8} tunes from underlying-event measurements'',} \textit{ Eur.
  Phys. J. C} \textbf{ 80} (2020) 4,
  \href{http://dx.doi.org/10.1140/epjc/s10052-019-7499-4}{\doi{10.1140/epjc/s10052-019-7499-4}},
  \href{http://www.arXiv.org/abs/1903.12179}{\texttt{arXiv:1903.12179}}.

\bibitem{Davidson:2010rw}
N.~Davidson\hrefCMSnoop {}{ { et~al.}, ``Universal interface of {TAUOLA}
  technical and physics documentation'',} \textit{ Comput. Phys. Commun.}
  \textbf{ 183} (2012) 821,
  \href{http://dx.doi.org/10.1016/j.cpc.2011.12.009}{\doi{10.1016/j.cpc.2011.12.009}},
  \href{http://www.arXiv.org/abs/1002.0543}{\texttt{arXiv:1002.0543}}.

\bibitem{Chrzaszcz:2016fte}
\hrefCMSnoop {}{M.~Chrzaszcz, T.~Przedzinski, Z.~Was, and J.~Zaremba,
  ``{TAUOLA} of $\tau$ lepton decays---framework for hadronic currents, matrix
  elements and anomalous decays'',} \textit{ Comput. Phys. Commun.} \textbf{
  232} (2018) 220,
  \href{http://dx.doi.org/10.1016/j.cpc.2018.05.017}{\doi{10.1016/j.cpc.2018.05.017}},
  \href{http://www.arXiv.org/abs/1609.04617}{\texttt{arXiv:1609.04617}}.

\bibitem{ParticleDataGroup:2022pth}
\hrefCMSnoop {}{{Particle Data Group}, S.~Navas { et~al.}, ``Review of particle
  physics'',} \textit{ Phys. Rev. D} \textbf{ 110} (2024) 030001,
  \href{http://dx.doi.org/10.1103/PhysRevD.110.030001}{\doi{10.1103/PhysRevD.110.030001}}.

\bibitem{LHCb:2019tea}
\hrefCMSnoop {}{{LHCb Collaboration}, ``Measurement of the {$\PB_\mathrm{c}^-$}
  meson production fraction and asymmetry in 7 and 13\,{TeV} {$\Pp\Pp$}
  collisions'',} \textit{ Phys. Rev. D} \textbf{ 100} (2019) 112006,
  \href{http://dx.doi.org/10.1103/PhysRevD.100.112006}{\doi{10.1103/PhysRevD.100.112006}},
  \href{http://www.arXiv.org/abs/1910.13404}{\texttt{arXiv:1910.13404}}.

\bibitem{LHCb:2013kwl}
\hrefCMSnoop {}{{LHCb Collaboration}, ``Observation of {$\PB^+_\mathrm{c}
  \rightarrow \JPsi \PD_\mathrm{s}^+$} and {$\PB^+_\mathrm{c} \rightarrow \JPsi
  \PD_\mathrm{s}^{*+}$} decays'',} \textit{ Phys. Rev. D} \textbf{ 87} (2013)
  112012,
  \href{http://dx.doi.org/10.1103/PhysRevD.87.112012}{\doi{10.1103/PhysRevD.87.112012}},
  \href{http://www.arXiv.org/abs/1304.4530}{\texttt{arXiv:1304.4530}}.
  [Erratum: \DOI{10.1103/PhysRevD.89.019901}].

\bibitem{ATLAS:2015jep}
\hrefCMSnoop {}{{ATLAS Collaboration}, ``Study of the {$\mathrm{B}_\mathrm{c}^+
  \rightarrow \JPsi \PD_\mathrm{s}^+$} and {$\PB_\mathrm{c}^+ \rightarrow \JPsi
  \PD_\mathrm{s}^{*+}$} decays with the {ATLAS} detector'',} \textit{ Eur.
  Phys. J. C} \textbf{ 76} (2016) 4,
  \href{http://dx.doi.org/10.1140/epjc/s10052-015-3743-8}{\doi{10.1140/epjc/s10052-015-3743-8}},
  \href{http://www.arXiv.org/abs/1507.07099}{\texttt{arXiv:1507.07099}}.

\bibitem{LHCb:2016djy}
\hrefCMSnoop {}{{LHCb Collaboration}, ``Observation of {$\text{B}_\text{c}^{+}
  \to \JPsi \mathrm{D}^{(*)} \PK^{(*)}$} decays'',} \textit{ Phys. Rev. D}
  \textbf{ 95} (2017) 032005,
  \href{http://dx.doi.org/10.1103/PhysRevD.95.032005}{\doi{10.1103/PhysRevD.95.032005}},
  \href{http://www.arXiv.org/abs/1612.07421}{\texttt{arXiv:1612.07421}}.

\bibitem{KISELEV2000473}
\hrefCMSnoop {}{V.~V. Kiselev, A.~K. Likhoded, and A.~I. Onishchenko,
  ``Semileptonic {$B_c$-meson} decays in sum rules of {QCD} and {NRQCD}'',}
  \textit{ Nucl. Phys. B} \textbf{ 569} (2000) 473,
  \href{http://dx.doi.org/10.1016/S0550-3213(99)00505-2}{\doi{10.1016/S0550-3213(99)00505-2}},
  \href{http://www.arXiv.org/abs/hep-ph/9905359}{\texttt{arXiv:hep-ph/9905359}}.

\bibitem{bgl:2019}
\hrefCMSnoop {}{T.~D. Cohen, H.~Lamm, and R.~F. Lebed, ``Precision
  model-independent bounds from global analysis of {$\PQb\to\PQc\ell\PGn$} form
  factors'',} \textit{ Phys. Rev. D} \textbf{ 100} (2019) 094503,
  \href{http://dx.doi.org/10.1103/PhysRevD.100.094503}{\doi{10.1103/PhysRevD.100.094503}},
  \href{http://www.arXiv.org/abs/1909.10691}{\texttt{arXiv:1909.10691}}.

\bibitem{hammer}
F.~U. Bernlochner\hrefCMSnoop {}{ { et~al.}, ``Das ist der {HAMMER}: Consistent
  new physics interpretations of semileptonic decays'',} \textit{ Eur. Phys. J.
  C} \textbf{ 80} (2020) 883,
  \href{http://dx.doi.org/10.1140/epjc/s10052-020-8304-0}{\doi{10.1140/epjc/s10052-020-8304-0}},
  \href{http://www.arXiv.org/abs/2002.00020}{\texttt{arXiv:2002.00020}}.

\bibitem{PhysRevD.107.052008}
\hrefCMSnoop {}{{HFLAV} Collaboration, ``Averages of b-hadron, c-hadron, and
  \ensuremath{\tau}-lepton properties as of 2021'',} \textit{ Phys. Rev. D}
  \textbf{ 107} (2023) 052008,
  \href{http://dx.doi.org/10.1103/PhysRevD.107.052008}{\doi{10.1103/PhysRevD.107.052008}},
  \href{http://www.arXiv.org/abs/2206.07501}{\texttt{arXiv:2206.07501}}.

\bibitem{Barberio:1993qi}
\hrefCMSnoop {}{E.~Barberio and Z.~W{\c a}s, ``{PHOTOS} --- a universal {Monte
  Carlo} for {QED} radiative corrections: version 2.0'',} \textit{ Comput.
  Phys. Commun.} \textbf{ 79} (1994) 291,
  \href{http://dx.doi.org/10.1016/0010-4655(94)90074-4}{\doi{10.1016/0010-4655(94)90074-4}}.

\bibitem{Agostinelli:2002hh}
\hrefCMSnoop {}{{GEANT4} Collaboration, ``{\GEANTfour}---a simulation
  toolkit'',} \textit{ Nucl. Instrum. Meth. A} \textbf{ 506} (2003) 250,
\href{http://dx.doi.org/10.1016/S0168-9002(03)01368-8}{\doi{10.1016/S0168-9002(03)01368-8}}.

\bibitem{Chen:2016btl}
\hrefCMSnoop {}{T.~Chen and C.~Guestrin, ``{XGBoost:} a scalable tree boosting
  system'',} 2016.
  \href{http://www.arXiv.org/abs/1603.02754}{\texttt{arXiv:1603.02754}}.
  \href{http://dx.doi.org/10.1145/2939672.2939785}{\doi{10.1145/2939672.2939785}}.

\bibitem{Phase1Pixel}
\hrefCMSnoop {}{{CMS Tracker Group}, W.~Adam { et~al.}, ``The {CMS} {Phase-1}
  pixel detector upgrade'',} \textit{ JINST} \textbf{ 16} (2021) P02027,
  \href{http://dx.doi.org/10.1088/1748-0221/16/02/P02027}{\doi{10.1088/1748-0221/16/02/P02027}},
  \href{http://www.arXiv.org/abs/2012.14304}{\texttt{arXiv:2012.14304}}.

\bibitem{ALEPH:2005qgp}
\hrefCMSnoop {}{{ALEPH} Collaboration, ``Branching ratios and spectral
  functions of tau decays: Final {ALEPH} measurements and physics
  implications'',} \textit{ Phys. Rept.} \textbf{ 421} (2005) 191,
  \href{http://dx.doi.org/10.1016/j.physrep.2005.06.007}{\doi{10.1016/j.physrep.2005.06.007}},
  \href{http://www.arXiv.org/abs/hep-ex/0506072}{\texttt{arXiv:hep-ex/0506072}}.

\bibitem{CMS:2024onh}
\hrefCMSnoop {}{{CMS Collaboration}, ``The {CMS} statistical analysis and
  combination tool: \textsc{Combine}'',} \textit{ Comput. Softw. Big Sci.}
  \textbf{ 8} (2024) 19,
  \href{http://dx.doi.org/10.1007/s41781-024-00121-4}{\doi{10.1007/s41781-024-00121-4}},
  \href{http://www.arXiv.org/abs/2404.06614}{\texttt{arXiv:2404.06614}}.

\bibitem{Sirunyan:2018nqx}
\hrefCMSnoop {}{{CMS Collaboration}, ``Measurement of the inelastic
  proton-proton cross section at $\sqrt{s}=13$\,{TeV}'',} \textit{ JHEP}
  \textbf{ 07} (2018) 161,
  \href{http://dx.doi.org/10.1007/JHEP07(2018)161}{\doi{10.1007/JHEP07(2018)161}},
\href{http://www.arXiv.org/abs/1802.02613}{\texttt{arXiv:1802.02613}}.

\bibitem{tracking}
\href {http://cds.cern.ch/record/2634940}{{CMS Collaboration}, ``Tracking {POG}
  results for pion efficiency with the {\PDst} meson using data from 2016 and
  2017'',} CMS Detector Performance Note CMS-DP-2018-050, 2018.

\bibitem{Fisher1992}
\hrefCMSnoop {}{R.~A. Fisher, ``Statistical Methods for Research Workers''},
  p.~66.
\newblock Springer New York, New York, NY, 1992.
\newblock
  \href{http://dx.doi.org/10.1007/978-1-4612-4380-9_6}{\doi{10.1007/978-1-4612-4380-9_6}}.

\end{thebibliography}\endgroup
\cleardoublepage \appendix\section{The CMS Collaboration \label{app:collab}}\begin{sloppypar}\hyphenpenalty=5000\widowpenalty=500\clubpenalty=5000\cmsinstitute{Yerevan Physics Institute, Yerevan, Armenia}
{\tolerance=6000
V.~Chekhovsky, A.~Hayrapetyan, V.~Makarenko\cmsorcid{0000-0002-8406-8605}, A.~Tumasyan\cmsAuthorMark{1}\cmsorcid{0009-0000-0684-6742}
\par}
\cmsinstitute{Institut f\"{u}r Hochenergiephysik, Vienna, Austria}
{\tolerance=6000
W.~Adam\cmsorcid{0000-0001-9099-4341}, J.W.~Andrejkovic, L.~Benato\cmsorcid{0000-0001-5135-7489}, T.~Bergauer\cmsorcid{0000-0002-5786-0293}, K.~Damanakis\cmsorcid{0000-0001-5389-2872}, M.~Dragicevic\cmsorcid{0000-0003-1967-6783}, C.~Giordano, P.S.~Hussain\cmsorcid{0000-0002-4825-5278}, M.~Jeitler\cmsAuthorMark{2}\cmsorcid{0000-0002-5141-9560}, N.~Krammer\cmsorcid{0000-0002-0548-0985}, A.~Li\cmsorcid{0000-0002-4547-116X}, D.~Liko\cmsorcid{0000-0002-3380-473X}, I.~Mikulec\cmsorcid{0000-0003-0385-2746}, J.~Schieck\cmsAuthorMark{2}\cmsorcid{0000-0002-1058-8093}, R.~Sch\"{o}fbeck\cmsAuthorMark{2}\cmsorcid{0000-0002-2332-8784}, D.~Schwarz\cmsorcid{0000-0002-3821-7331}, M.~Shooshtari\cmsorcid{0009-0004-8882-4887}, M.~Sonawane\cmsorcid{0000-0003-0510-7010}, W.~Waltenberger\cmsorcid{0000-0002-6215-7228}, C.-E.~Wulz\cmsAuthorMark{2}\cmsorcid{0000-0001-9226-5812}
% ...
\par}
\vskip\cmsinstskip
\dag:~Deceased\\
$^{1}$Also at Yerevan State University, Yerevan, Armenia\\
$^{2}$Also at TU Wien, Vienna, Austria\\
% ...
\end{sloppypar}
\end{document}